\documentclass[aps,prl,twocolumn,superscriptaddress]{revtex4}
\usepackage[utf8]{inputenc}
\usepackage{amsmath}
\usepackage{amssymb}
\usepackage{xcolor}
\usepackage{graphicx}

\newcommand{\bk}{\mathbf{k}}
\newcommand{\bkp}{\mathbf{k'}}

\newcommand{\bq}{\mathbf{q}}

\def\dz2{d$_{\text{z}^2}$}

\def\dx2y2{d$_{\text{x}^2\text{y}^2}$}
\def\G0W0{G$_0$W$_0$}
\def\scGW0{scGW$_0$}

\def\mos{MoS$_2$}

\begin{document}

\author{A. Steinhoff}
\affiliation{Institut f\"ur Theoretische Physik, Universit\"at Bremen, P.O. Box 330 440, 28334 Bremen, Germany}
\author{M. Florian}
\affiliation{Institut f\"ur Theoretische Physik, Universit\"at Bremen, P.O. Box 330 440, 28334 Bremen, Germany}
\author{M. R\"osner}
\affiliation{Institut f\"ur Theoretische Physik, Universit\"at Bremen, P.O. Box 330 440, 28334 Bremen, Germany}
\affiliation{Bremen Center for Computational Materials Science, Universit\"at Bremen, 28334 Bremen, Germany}
\author{M. Lorke}
\affiliation{Institut f\"ur Theoretische Physik, Universit\"at Bremen, P.O. Box 330 440, 28334 Bremen, Germany}
\author{T.O. Wehling}
\affiliation{Institut f\"ur Theoretische Physik, Universit\"at Bremen, P.O. Box 330 440, 28334 Bremen, Germany}
\affiliation{Bremen Center for Computational Materials Science, Universit\"at Bremen, 28334 Bremen, Germany}
\author{C. Gies}
\affiliation{Institut f\"ur Theoretische Physik, Universit\"at Bremen, P.O. Box 330 440, 28334 Bremen, Germany}
\author{F. Jahnke}
\affiliation{Institut f\"ur Theoretische Physik, Universit\"at Bremen, P.O. Box 330 440, 28334 Bremen, Germany}
\email{asteinhoff@itp.uni-bremen.de}

\title{Nonequilibrium Carrier Dynamics in Transition Metal Dichalcogenide Semiconductors}

\maketitle

\textbf{When exploring new materials for their potential in (opto)electronic device applications, it is important to understand the role of various carrier interaction and scattering processes. Research on transition metal dichalcogenide (TMD) semiconductors has recently progressed towards the realisation of working devices, which involve light-emitting diodes~\cite{withers_light-emitting_2015}, nanocavity lasers~\cite{wu_monolayer_2015}, and single-photon emitters~\cite{he_single_2015,chakraborty_voltage-controlled_2015}. In these two-dimensional atomically thin semiconductors, the Coulomb interaction is known to be much stronger than in quantum wells of conventional semiconductors like GaAs, as witnessed by the 50 times larger exciton binding energy~\cite{ugeda_giant_2014}. The question arises, whether this directly translates into equivalently faster carrier-carrier Coulomb scattering of excited carriers. Here we show that a combination of ab-initio band-structure and many-body theory predicts carrier 
relaxation on a 50-fs time scale, which is less than an order of magnitude faster than in quantum wells. These scattering times compete with the recently reported sub-ps exciton recombination times~\cite{poellmann_resonant_2015}, thus making it harder to achieve population inversion and lasing.}

In the past, for conventional semiconductors, considerable experimental and theoretical activities have been devoted to identify and characterize the intrinsic interaction processes. Only on these grounds, it was possible to
reveal the true device operation potential of a material. For the new TMD semiconductors, experiments uncovering carrier relaxation processes in connection with optical properties are at a very early stage. Among the experiments are time-resolved differential transmission measurements demonstrating rapid thermalisation and cooling of excited carriers~\cite{nie_ultrafast_2014}, ultrafast differential reflection changes due to excited carriers~\cite{chernikov_population_2015}, and excitation-density and temperature dependent homogeneous linewidth studies~\cite{moody_intrinsic_2015}.
While these experiments focus on below-band-gap excitation of excitons, the device-relevant case of elevated carrier densities with large excess energy, exhibiting a different relaxation dynamics, has not been addressed yet.
In the present work, we study the relaxation of optically excited carriers with large excess energy under the influence of carrier-carrier Coulomb interaction in a freestanding monolayer of \mos\ to quantify the material limits of carrier scattering rates. The results are compared to carrier dynamics predicted for quasi-resonant optical pumping of excitons. For this purpose, we determine the evolution of the initial nonequilibrium carrier distributions and their relaxation across the whole Brillouin zone for the different excitation conditions, which are shown to strongly influence the scattering efficiency. Within our approach, we are able to address separately the dynamics of electrons and holes that is governed by the characteristic valley structures of the respective bands. 
This type of investigations critically depends on a realistic description of the band structure and on the material-specific enhancement of the Coulomb interaction. The latter needs to include the particular quasi-two-dimensional nature of the atomically thin material as well as the reduced dielectric background screening of the Coulomb interaction as discussed in~\cite{steinhoff_influence_2014}. In our calculations, the role of these ground-state properties is complemented by screening and many-particle band structure renormalisations due to dynamically changing distributions of excited carriers.

\paragraph{Carrier Dynamics After Femtosecond Pulse Excitation.}

\begin{figure*}[t!]
  \begin{center}
    \includegraphics[width=\textwidth]{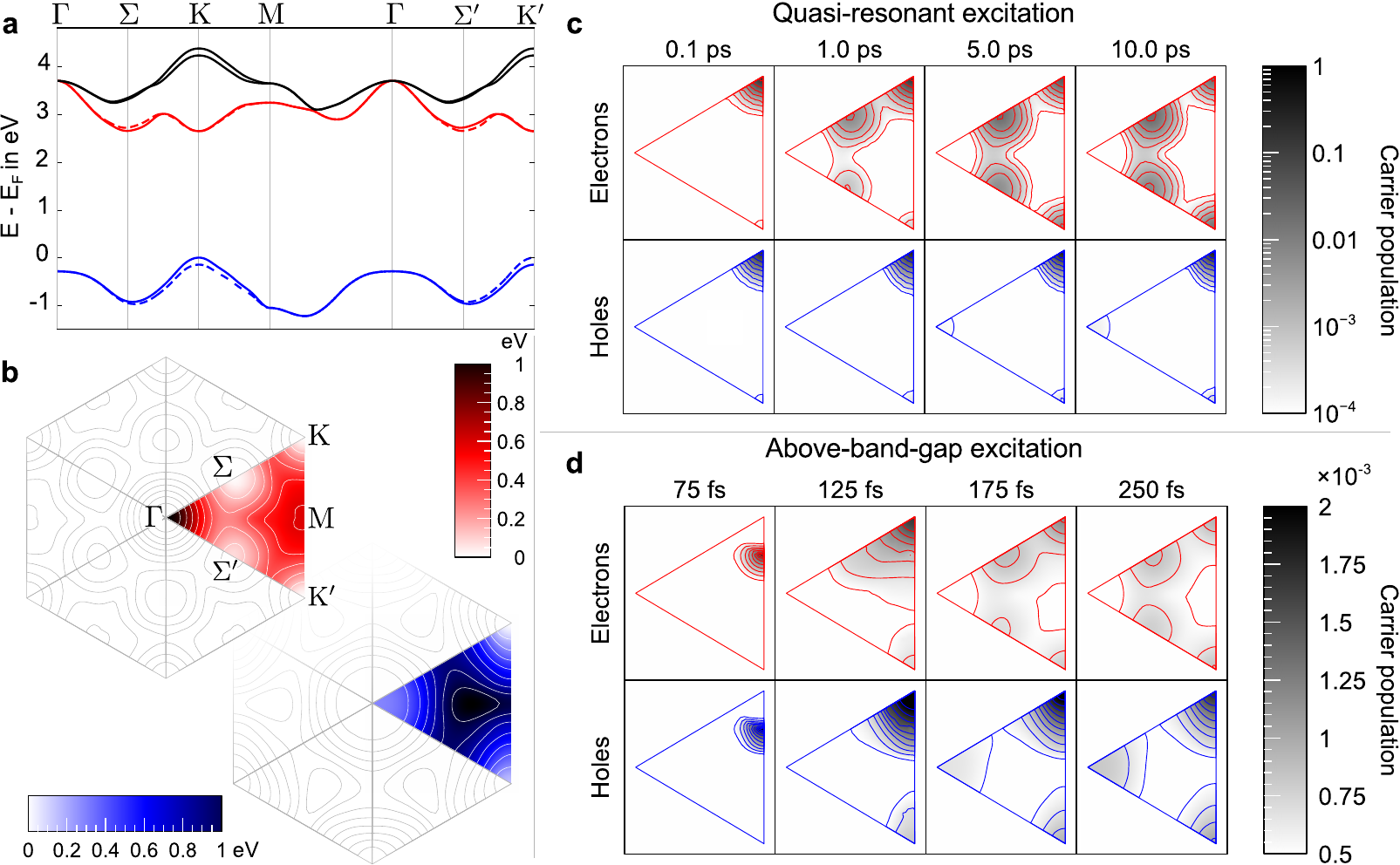}
    \caption{\textbf{Microscopic characterisation of carrier dynamics.} \textbf{a,} Band structure of monolayer \mos\ as obtained from a \G0W0 calculation including spin-orbit interaction. The relevant conduction and valence bands are marked in colour, while relevant bands with positive and negative z-direction of spin are represented by solid and dashed lines, respectively. \textbf{b,} Band structure of electrons and holes with positive z-direction of spin over the full Brillouin zone (BZ) corresponding to the curves in the top figure with gap energy subtracted. The irreducible part of the BZ is highlighted and high-symmetry points are shown, where band-structure valleys are depicted in white. 
    \textbf{c,} Time evolution of the carrier populations with positive z-direction of spin on the irreducible part of the BZ after quasi-resonant excitation. \textbf{d,} Same as \textbf{c} after above-band-gap excitation. For a discussion of the other spin direction, see the Supplementary Material. The optical excitation pulse has its peak amplitude at $75$\,fs. }
    \label{fig1}
  \end{center}
\end{figure*}
In the following, we consider two different excitation scenarios that directly relate to most optical experiments performed on TMD semiconductors. Quasi-resonant excitation of the monolayer \emph{below} the single-particle band gap is facilitated by the existence of excitonic states, which can be coherently driven by a resonant optical pulse. Excitation \emph{above} the single-particle band gap by optical pulses with shorter wavelengths on the other hand generates an electron-hole plasma. We show that although both scenarios result in ultrafast carrier dynamics throughout the Brillouin zone, the observed time scales are different. To this end, we solve the kinetic equations discussed in the Methods section, which include optical transitions between and carrier dynamics in the two highest valence bands and the two lowest conduction bands. The latter are highlighted in Fig.~\ref{fig1}a. The theory also includes the interband electron-hole Coulomb interaction as well as the intraband electron-electron and hole-
hole interaction.

Optical excitation is described by a Gaussian laser pulse with a temporal width of $50\,$fs, corresponding to a spectral width of $73\,$meV, and applied perpendicular to the freestanding layer assuming circular light polarisation. For the first scenario of quasi-resonant excitation the wavelength is chosen so that it partly overlaps with the A-exciton resonance (detuning $60\,$meV above A exciton), while above-band-gap excitation takes place 600 meV above the single-particle gap (corresponding to a 405 nm wavelength). The homogeneous linewidth of the optical transition is described in Eq.~(\ref{eq:psik}) by a phenomenological dephasing $\gamma=10$ meV. In both excitation scenarios, the deposited energy per pulse is chosen to create a carrier density of $3\times10^{12}$ cm$^{-2}$.

\begin{figure*}[t!]
 \begin{center}
   \includegraphics[width=\textwidth]{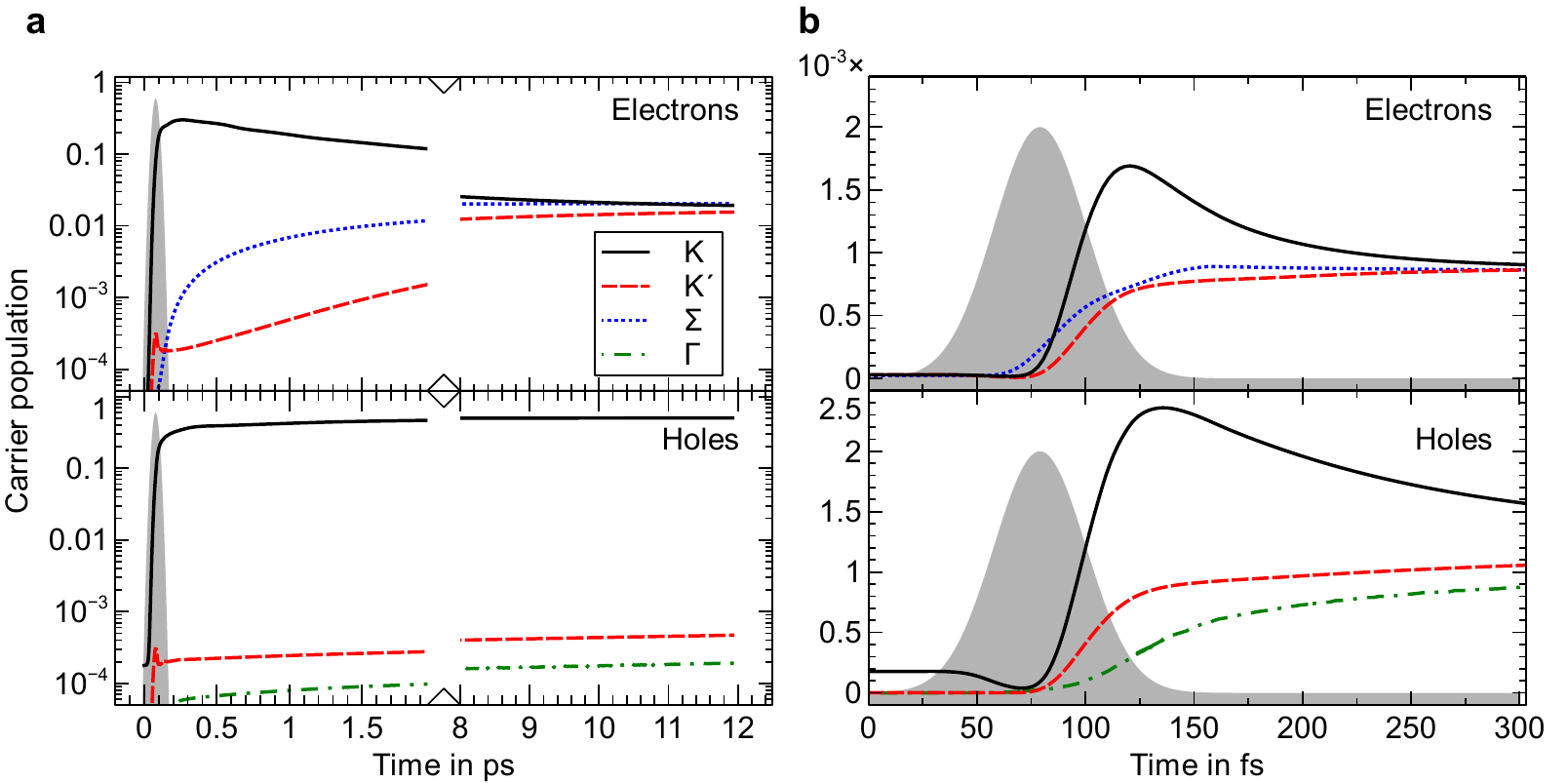}
   \caption{\textbf{Time evolution of carrier populations at high symmetry points.} \textbf{a,} Quasi-resonant excitation of electrons and holes. \textbf{b,} Above-band-gap excitation of electrons and holes. The shape of the laser pulse is shown as a shaded grey area.}
   \label{fig2}
 \end{center}
\end{figure*}

The time evolution of the electron and hole distribution over the irreducible part of the Brillouin zone is shown in Fig.~\ref{fig1}c and d for quasi-resonant and above-band-gap excitation, respectively. In the case of quasi-resonant excitation, the coherent optical field creates a polarisation in the system due to its spectral overlap with the excitonic resonance. 
The result is a population of carriers at the respective $\bk$-points where the bound-state wave functions reside. For the A exciton, this is mainly in the K valleys of the Brillouin zone \cite{hill_observation_2015,steinhoff_influence_2014,qiu_optical_2013}, as can be seen in Fig.~\ref{fig1}c at time 0.1 ps. Subsequently, carrier populations are distributed among the different valleys in the band structure on a ps time scale, as depicted in Fig.~\ref{fig2}a. Electrons are first drawn from the K to the $\Sigma$ valley and later also to $\Sigma$' and K', which are far off in $\bk$-space. After about $10$ ps, K, K' and $\Sigma$ valley populations of electrons are almost equal, while the $\Sigma$' valley is less populated due to its separation by spin-orbit splitting.

The main feature observed in the hole distribution is a narrowing as it approaches a Fermi distribution function, whereas only a small fraction of holes is lost to K' and $\Gamma$. The energy separation between K and K' valleys, which is determined by the spin-orbit interaction, is large enough to prevent efficient inter-valley transfer of holes. As a result of this energy barrier, even the fast carrier-carrier Coulomb scattering (which itself conserves the spin) can redistribute only a small fraction of holes with a given spin polarisation between K and K' valleys. This and the fact that spin-flip processes occur on a much longer ns time scale is what enables spintronics applications based on monolayer TMDs \cite{mai_many-body_2014,yang_long-lived_2015}.

By exciting the monolayer 600 meV above the single-particle gap, an electron-hole plasma is created in those $\bk$-states overlapping energetically with the optical pulse according to the momentum-dependent dipole coupling. This leads to an excess energy of carriers corresponding to a temperature of about 10000 K. The resulting carrier distribution in the Brillouin zone is shown in Fig.~\ref{fig1}d after 75 fs, when the optical pulse has reached its peak value. We find that already 50 fs later, electrons and holes have relaxed to their corresponding valleys in the band structure, see also Fig.~\ref{fig2}b, which shows that carrier relaxation takes place on the same time scale as the ultrashort optical pulse. 
In graphene as another two-dimensional material, longer relaxation times of about 100 fs have been found \cite{malic_microscopic_2011}. One reason for this difference is the orientational preference of carrier scattering in graphene that strongly favors co-linear scattering, which does not exist in TMDs. After the ultrafast initial relaxation, carriers redistribute slightly among the valleys as electrons and holes equilibrate to a common temperature. This process is determined by the slower dynamics of the hole relaxation. The same tendency was found in \cite{nie_ultrafast_2014} for few-layer \mos. The origin for carrier relaxation from high-energy states being faster than inter-valley scattering is the better availability of final states for the assisting scattering process. Additionally, inter-valley scattering requires larger momentum transfer, for which the involved Coulomb matrix elements are smaller.

In comparison, both excitation schemes lead to qualitatively different carrier distributions in the Brillouin zone: By exciting above the gap, a hot electron-hole plasma in the band structure valleys is created on a sub-100 fs time scale, while quasi-resonant excitation below the gap leads to a strong electron and hole population in the K valley and carriers are lost to other valleys only on a slower time scale of several ps.

The final quasi-equilibrium carrier distribution depends on the relative energetic positions of the valleys in each band. When considering energy renormalisations of the band structure due to many-body interaction of excited carriers, we find that the valleys exhibit different energy shifts. The corresponding results are provided in the Supplementary Material. Similar to modifications of the band structure by strain \cite{conley_bandgap_2013,steinhoff_efficient_2015}, the excitation-induced shifts after quasi-resonant excitation mediate a transition from direct to indirect semiconductor during the inter-valley scattering. Moreover, the renormalisations cause a band-gap shrinkage of about 200 meV for the excitation parameters used here.
Another effect that is important for the carrier dynamics besides many-particle renormalisations is the screening of Coulomb interaction due to excited carriers, which strongly depends on the carrier distribution in $\bk$-space and hence on the excitation scenario. As shown in the Supplementary Material, this affects the carrier-carrier scattering efficiency and is a strong contribution to the observed different time scales for the relaxation dynamics.

In conclusion, we have studied the excited-carrier dynamics due to carrier-carrier Coulomb scattering in TMD semiconductor systems on the basis of a material-realistic ab-initio description of the electronic band structure and Coulomb interaction in combination with a many-body theory of carrier dynamics. The efficiency of carrier redistribution throughout the Brillouin zone is shown to be determined by an interplay of carrier scattering and time-dependent many-particle renormalisations of the band structure as well as screening of the Coulomb interaction. As a result, thermalisation on a $50\,$fs timescale following above-band-gap excitation is possible due to efficient carrier-carrier scattering in the two-dimensional material. We report significantly slower relaxation on a $5\,$ps timescale for quasi-resonant excitation below the band gap. In this case, carriers are mainly generated in the K valleys, from where further redistribution requires inter-valley scattering with a large momentum transfer.
In conventional semiconductors based on GaAs, carrier-carrier scattering leads to a redistribution of excited carriers towards Fermi-Dirac-functions on a timescale of 100 fs. \cite{knox_femtosecond_1988, knox_femtosecond_1986, shah_ultrafast_1999} Having the huge exciton binding energy in two-dimensional TMDs in mind, the carrier scattering efficiency in these materials appears to be smaller than expected. A comparison to recently reported exciton recombination times of 150 fs~\cite{poellmann_resonant_2015} leads us to the conclusion that the ratio of both time scales is less favourable for achieving population inversion and laser operation than in conventional semiconductors.

\paragraph{Methods.}
We combine material-realistic ab-initio methods with a many-body theory of the carrier dynamics. Band structures and Coulomb matrix elements from a G$_0$W$_0$ calculation are used as a basis for excited-carrier interaction~\cite{steinhoff_influence_2014}.

To describe the dynamics of excited carriers induced by pulsed optical excitation of the TMD system, we solve coupled equations of motion for electron and hole populations $f_{\bk}^{\lambda}(t)=\big<a^{\dagger,\lambda}_{\bk} a^{\phantom\dagger \lambda}_{\bk}\big>$ and microscopic interband polarisations $\psi_{\bk}^{he}(t)=\big<a^{h}_{\bk} a^{e}_{\bk}\big>$:

\begin{align}
\frac{\mathrm{d}}{\mathrm{dt}}\psi_{\bk}^{he}(t) & =-\frac{i}{\hbar}\left( \widetilde{\varepsilon}_{\bk}^{h}(t) + \widetilde{\varepsilon}_{\bk}^{e}(t)-i\gamma\right) \label{eq:psik} \\
\quad +i\Omega_{\bk}^{he}(t)&\left( 1-f_{\bk}^{h}(t)-f_{\bk}^{e}(t)\right)\,, \notag \\
\frac{\mathrm{d}}{\mathrm{dt}}f_{\bk}^{e/h}(t) & = 2\,\mathrm{Im}\left((\Omega_{\bk}^{he}(t))^* \psi_{\bk}^{he}(t) \right) \label{eq:fk} \\
+\mathrm{S}^{e/h,\mathrm{in}}_{\bk}(t)&(1-f_{\bk}^{e/h}(t))-\mathrm{S}^{e/h,\mathrm{out}}_{\bk}(t)f_{\bk}^{e/h}(t)\,. \notag
\end{align}

The renormalised single-particle energies 
\begin{equation}
 \begin{split} 
 \widetilde{\varepsilon}_{\bk}^{\lambda}(t) = \varepsilon_{\bk}^{\lambda}
 &+\frac{1}{A}\sum_{\bk',\lambda'}V^{\lambda\lambda'\lambda'\lambda}_{\bk\bkp\bkp\bk}f_{\bkp}^{\lambda}(t) \\
 &-\frac{1}{A}\sum_{\bk',\lambda'}V^{\lambda\lambda'\lambda\lambda'}_{\bk\bkp\bk\bkp}f_{\bkp}^{\lambda}(t) 
 \label{eq:HF}
 \end{split}
\end{equation}
contain the free-carrier energies given by the band structure, as well as Hartree-Fock energy corrections due to the excited-carrier populations $f_{\bk}$. 
Owing to the strong impact of Coulomb effects in atomically thin systems, band structure $\varepsilon_{\bk}^{\lambda}$ and Coulomb matrix elements $V^{\lambda\lambda'\lambda\lambda'}_{\bk\bkp\bk\bkp}$ must be chosen as accurately as possible for the given TMD material system. We obtain both from ab-initio calculations on the G$_0$W$_0$-level using the approach introduced in~\cite{steinhoff_influence_2014,steinhoff_efficient_2015}. In the calculation of these matrix elements dielectric screening due to charge carriers in the ground state of the system is already included.

The interband polarisation is driven by the renormalised Rabi energy,
\begin{equation}
    \hbar\Omega_{\bk}^{he}(t)=E(t) d_{\bk}^{he}+\frac{1}{A}\sum_{\bk'}V_{\bk\bk'\bk\bk'}^{ehhe}\psi_{\bk'}^{he}(t) ~,
    \label{eq:rabi}
\end{equation}
accounting for the reaction of the electron-hole system on the applied optical field $E(t)$, mediated by the interband dipole coupling matrix element $d^{he}_{\bk}$, as well as the electron-hole Coulomb interaction~\cite{steinhoff_influence_2014}.

Eq.~(\ref{eq:psik}) together with Eq.~(\ref{eq:rabi}) represent a two-particle Schr\"odinger equation for the relative motion of optically driven electron-hole pairs. As a result, optically induced interband transitions, described by $\psi_{\bk}$, represent \emph{pair excitations under the influence of Coulomb interaction}, so that Eq~(\ref{eq:rabi}) contains transitions of both excitonic bound and scattering states. The corresponding population of excited carriers in connection with the optical excitation is described by Eq.~(\ref{eq:fk}).  

On the level of a Boltzmann equation \cite{schafer_semiconductor_2002}, dynamical changes of the single-particle populations are described by collision terms accounting for in- and
out-scattering contributions for each single-particle state $\bk$ in the Brillouin zone with corresponding rates $\mathrm{S}^{e/h,\mathrm{in}}_{\bk}(t)$ and $\mathrm{S}^{e/h,\mathrm{out}}_{\bk}(t)$, respectively.
For the carrier-carrier scattering due to Coulomb interaction, one obtains in second-order Born and Markov approximation

\begin{equation}
\begin{split}
\mathrm{S}^{e,\mathrm{in}}_{\bk}(t) & = \frac{2\pi}{\hbar}\sum_{\bkp\bq}\sum_{\lambda} \\
\times\delta(&\widetilde{\varepsilon}_{\bk}^{e}(t) + \widetilde{\varepsilon}_{\bkp-\bq}^{\lambda}(t)- \widetilde{\varepsilon}_{\bkp}^{\lambda}(t)-\widetilde{\varepsilon}_{\bk+\bq}^{e}(t) ) \\
\times\Big( | & W_{\bk,\bkp-\bq,\bkp,\bk+\bq}^{e\lambda\lambda e}(t)|^2 \\
- & W_{\bk,\bkp-\bq,\bkp,\bk+\bq}^{e\lambda\lambda e}(t)(W_{\bk,\bkp-\bq,\bk+\bq,\bkp}^{e\lambda e\lambda}(t))^* \Big) \\
&\times(1-f_{\bkp-\bq}^{\lambda}(t))f_{\bkp}^{\lambda}(t)f_{\bk+\bq}^{e}(t)\,, \\
\mathrm{S}^{e,\mathrm{out}}_{\bk}(t) & =  \frac{2\pi}{\hbar}\sum_{\bkp\bq}\sum_{\lambda} \\
\times\delta(&\widetilde{\varepsilon}_{\bk}^{e}(t) + \widetilde{\varepsilon}_{\bkp-\bq}^{\lambda}(t)- \widetilde{\varepsilon}_{\bkp}^{\lambda}(t)-\widetilde{\varepsilon}_{\bk+\bq}^{e}(t) ) \\
\times\Big( | & W_{\bk,\bkp-\bq,\bkp,\bk+\bq}^{e\lambda\lambda e}(t)|^2 \\
- & W_{\bk,\bkp-\bq,\bkp,\bk+\bq}^{e\lambda\lambda e}(t)(W_{\bk,\bkp-\bq,\bk+\bq,\bkp}^{e\lambda e\lambda}(t))^* \Big) \\
&\times f_{\bkp-\bq}^{\lambda}(t)(1-f_{\bkp}^{\lambda}(t)(1-f_{\bk+\bq}^{e}(t))\,,
\label{eq:scattrate}
\end{split}
\end{equation}
as well as analogue expressions for hole population functions.

Two additional approximations in Eq.~(\ref{eq:scattrate}) need to be addressed. The delta-functions describing energy conservation are formulated with energies $\widetilde{\varepsilon}_{\bk}(t)$ given by the ground-state G$_0$W$_0$ band structure, which is renormalised by time-dependent Hartree-Fock shifts induced by the excited carriers. 
In a derivation of Eq.~(\ref{eq:scattrate}) based on nonequilibrium Green's functions, this is the result of a quasi-particle approximation, which is introduced in the spirit of perturbation theory. 
Furthermore, the scattering rates contain statically screened Coulomb matrix elements. The nonequilibrium Green's functions technique naturally includes the screening of the interaction as part of the diagram summation leading to the second-order Born approximation. Essentially, the form of Eq.~(\ref{eq:scattrate}) is obtained by including all terms of quadratic order in the screened Coulomb interaction
\begin{equation}
    \begin{split}
        W_{\bk,\bkp-\bq,\bkp,\bk+\bq}^{e\lambda\lambda e}(t)=
        \varepsilon^{-1,\textrm{excited}}_{\bq}(t) V_{\bk,\bkp-\bq,\bkp,\bk+\bq}^{e\lambda\lambda e} ~.
        \label{eq:coulme}
    \end{split}
\end{equation}
We treat the screening $\varepsilon^{-1,\textrm{excited}}_{\bq}$, which results from the excited carriers in addition to the dielectric screening from the material, in terms of the Lindhard formula. Details on the numerical solution of the coupled equations and the screening are given in the Supplementary Material.

\paragraph{Acknowledgement}
This work has been supported by the Deutsche Forschungsgemeinschaft. The authors acknowledge resources for computational time at the HLRN (Hannover/Berlin).

\appendix

\section{Supplementary Material.}

\subsection{Carrier Spins.}

Due to the mirror symmetry of monolayer transition metal dichalcogenides with respect to the x-y plane (where z is the growth direction), each electronic band has a well-defined z component of spin, even in the presence of spin-orbit interaction. The latter causes a significant splitting of bands, especially at the K point of the valence band and at the $\Sigma$ point of the conduction band. \cite{liu_three-band_2013} There are no optical selection rules with respect to carrier spin but only to the valley degree of freedom. Hence circularly polarized light excites electrons and holes in both spin directions.

For the discussion of results in the main text we restrict ourselves to the time evolution of carriers with positive z-direction of spin, corresponding to the A-exciton resonance, as results are similar for the other spin polarisation. Modifications stem from the different energetic position of the B-exciton resonance and the spin-down band structure relative to the optical pulse energy. Note that although carrier-carrier-Coulomb interaction is not able to flip carrier spins, excited carriers in different spin-subsystems influence each other significantly via the electrostatic Hartree interaction, by taking up excess energy in carrier-carrier-scattering processes and by contributing to the screening of Coulomb interaction.

\subsection{Hartree-Fock Renormalisations.}

\begin{figure*}[t!]
 \begin{center}
   \includegraphics[width=\linewidth]{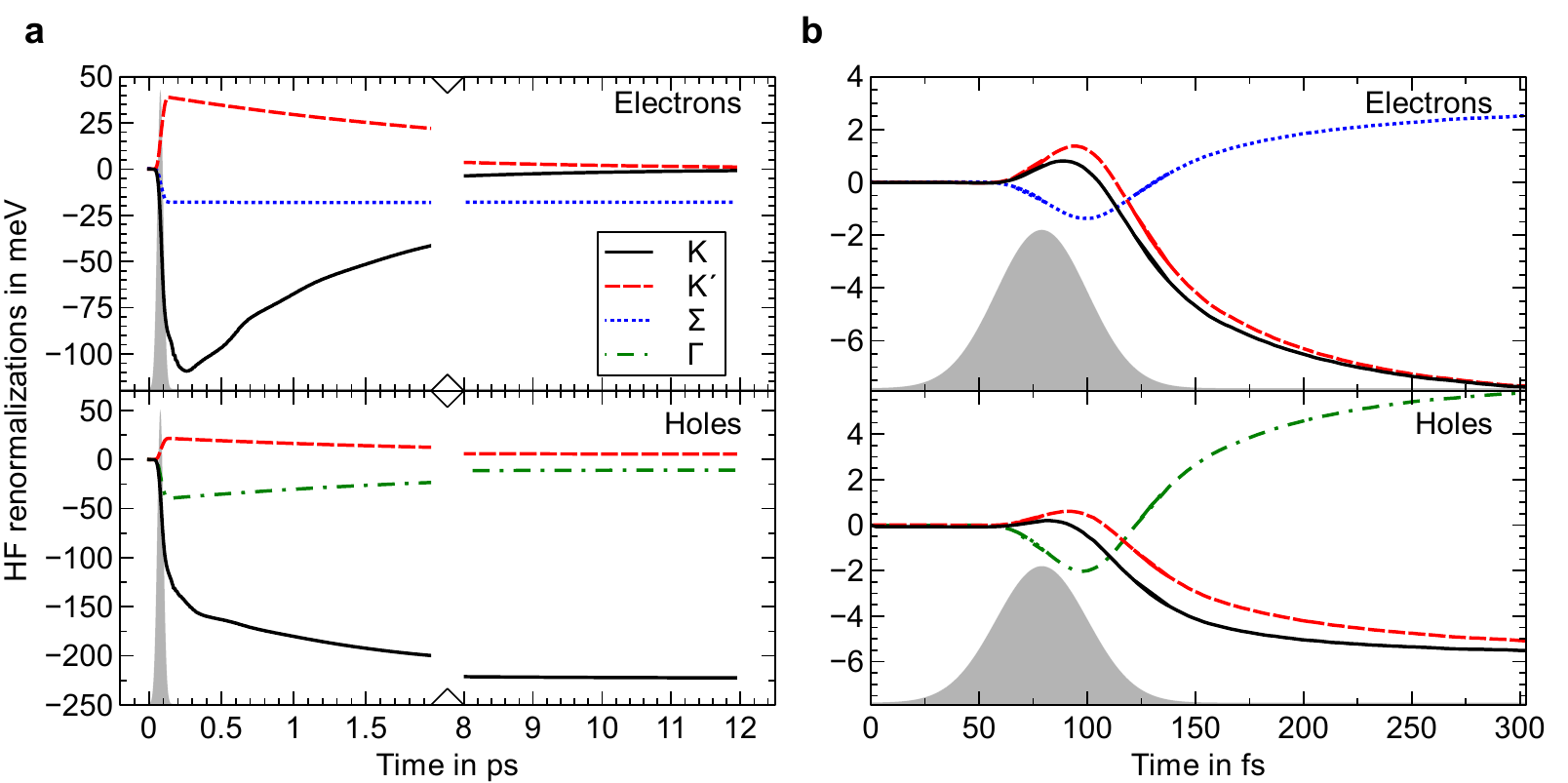}
   \caption{\textbf{Time-dependent Hartree-Fock renormalisations at high symmetry points of the Brillouin zone.} \textbf{a,} Quasi-resonant excitation of electrons and holes. \textbf{b,} Above-band-gap excitation of electrons and holes. The shape of the laser pulse is shown as a shaded grey area.}
   \label{figS1}
 \end{center}
\end{figure*}
In Fig.~\ref{figS1}, we provide results for the time-dependent band structure renormalisations due to Hartree-Fock interaction, caused by the excited carriers, which enter the carrier-scattering rates in Eq.~(5). For the quasi-resonant excitation scenario presented in Fig.~\ref{figS1}a, we find that both electron and hole K valleys are significantly lowered during the excitation pulse by exchange interaction. The lowering of hole energies in the K valley proceeds during carrier relaxation, as a Fermi distribution forms in the dynamically renormalized band structure until self-consistence of energies and populations is reached. This leads to a single-particle-gap shrinkage of about 200 meV and a reduced effective mass of the valence band K valley. On the other hand, the exchange shift of electrons in the K valley is reduced during inter-valley scattering, as electrons are lost to the $\Sigma$ valley. The K'-valley energies of electrons and holes as well as the $\Gamma$-valley energy are initially shifted due 
to Hartree interaction with the carriers in the K valleys. At later stages of the time evolution, these shifts are reduced, as carriers redistribute among the valleys. The $\Sigma$ valley experiences an almost constant red shift due to Hartree and exchange interaction, which is in equilibrium situation (reached at later times) larger than the K-valley shift. In combination with the ground-state band structure, the net result is a shift of the $\Sigma$ valley below K. Thus the many-particle renormalisations cause a transformation from direct to indirect semiconductor.

For above-band-gap excitation, dynamical Hartree-Fock shifts of the band-structure valleys are shown in Fig.~\ref{figS1}b. As carriers are not generated in the valleys but in the center of the Brillouin zone, the initially induced renormalisations in the valleys are small. During carrier relaxation, they become larger, but remain in the meV region due to the much broader population functions compared to quasi-resonant excitation.

\subsection{Excited-carrier Screening.}

\begin{figure*}[t!]
  \begin{center}
    \includegraphics[width=\textwidth]{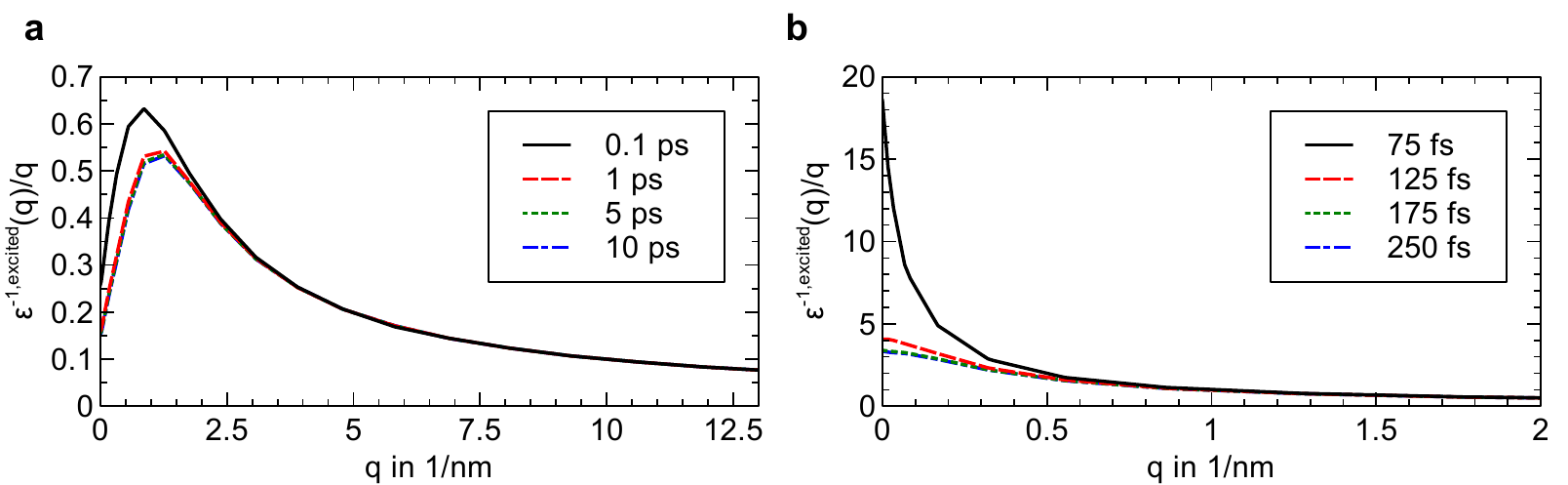}
    \caption{\textbf{Momentum dependence of Coulomb potential screened by excited carriers at different times of the relaxation dynamics as shown in Fig.1.} \textbf{a,} Quasi-resonant excitation. \textbf{b,} Above-band-gap excitation. The momentum $\bq$ is taken on a contour from $\Gamma$ point to $K$ point. Note the different momentum range in both cases.}
    \label{figS2}
  \end{center}
\end{figure*}

We describe the screening of the Coulomb interaction in TMD systems, which is caused by excited carriers, by a static (frequency-independent) Lindhard formula involving the dynamically changing carrier populations $f_{\bk}^{\lambda}(t)$ from the full Brillouin zone:

\begin{equation} 
\begin{split}
  & \varepsilon^{-1,\textrm{excited}}_{\bq}(t) = \\ & 1-\sum_{\lambda}
\int \frac{d^2\bk}{(2\pi)^2} V_{\bk,\bk-\bq,\bk,\bk-\bq}^{\lambda\lambda\lambda\lambda}
\frac{f_{\bk-\bq}^{\lambda}(t)-f_{\bk}^{\lambda}(t)}{\varepsilon_{\bk-\bq}^{\lambda}-\varepsilon_{\bk}^{\lambda}+i\gamma}~.
 \label{eq:lindhard}
 \end{split}
\end{equation}

The ground-state band structure is used in the denominator. For numerical stability a small phenomenological damping $\gamma=10^{-2}$ meV and a small residual electron and hole density of
$10^9$ cm$^{-2}$ is assumed to populate the system before the optical excitation to enable finite screening at all times. 

In Fig.~\ref{figS2}, we show the scaled inverse dielectric function at different times (corresponding to Fig.1 in the main text) for above-gap and quasi-resonant optical excitation as discussed in the main text. We find that for above-gap excitation, screening is relatively weak at early times, as carrier populations are generated in $\bk$-space far from the band-structure valleys. This behaviour is characteristic for two-dimensional systems and can already be estimated in the long-wavelength (small $q$) limit in an effective-mass model, where screening is only caused by the carrier population in the valleys. \cite{haug_quantum_1993} As carriers are scattered into the valleys the screening efficiency increases.

\subsection{Influence of Phonons.}

In the present work we only consider carrier dynamics due to carrier-carrier scattering and neglect the effects of phonons on carrier-scattering times and carrier temperature. In the early stage of the relaxation dynamics, this assumption is supported by the fact that the monolayer phonons will be heated themselves by the interaction with the excited carriers, which reduces their carrier-scattering and cooling efficiency. On a typical time scale of 1 ps \cite{nie_ultrafast_2014}, the monolayer phonons, and with them the carriers, are expected to thermalize to the ambient temperature due to interaction with the substrate. For above-band-gap excitation, this is expected not to interfere with the faster dynamics due to carrier-carrier scattering discussed here. On the other hand, in few-layer \mos\, carrier-phonon interaction has been identified as a source for efficient inter-valley scattering on the sub-ps time scale. In \cite{nie_ultrafast_2015}, this is found to be the most efficient mechanism for inter-valley scattering. The result is consistent with our findings of carrier-carrier-mediated inter-valley scattering taking place on a longer time scale and hence being the less important mechanism here. In the end, the role of phonons will be much more dependent on the external heat management of the TMD layers in comparison to conventional semiconductors.

\subsection{Numerical Methods.}

The ab-initio calculations are performed as described in detail in \cite{steinhoff_influence_2014}. The Brillouin-zone integral in Eq.~(\ref{eq:lindhard}) is evaluated by means of adaptive cubature over a triangular region using the DCUTRI algorithm. \cite{berntsen_algorithm_1992} For this purpose, the population functions and band structure under the $\bk$-integral are interpolated with a cubic Shepard method. \cite{renka_algorithm_1999}

The equations of motion (1) and (2) in combination with equations (3) to (6) are integrated in time using an adaptive predictor-corrector method. \cite{brown_vode:_1989} We use a Monkhorst-Pack grid to sample the Brillouin zone. We find that in the case of quasi-resonant optical excitation, 48x48x1 grid points are sufficient to converge the carrier-carrier scattering results, while in the case of above-gap excitation, the carrier distribution in $\bk$-space generated by the optical pulse is better resolved by a 60x60x1-grid. Due to symmetry, the equations of motion have to be solved for states $\bk$ from the irreducible Brillouin zone only, while summation over $\bk'$ and $\bq$ describing interaction processes has to run over the full Brillouin zone. 

To evaluate the energy-conserving Dirac-Delta functions in the carrier-carrier scattering integrals (Eq.~(5)), we successively replace the Delta functions by appropriately normalised Gaussians with decreasing width and using a Romberg extrapolation to obtain the scattering integrals in the limit of zero width.

\bibliography{2015_MoS2_Coulomb_scattering}

\begin{thebibliography}{27}
\expandafter\ifx\csname natexlab\endcsname\relax\def\natexlab#1{#1}\fi
\expandafter\ifx\csname bibnamefont\endcsname\relax
  \def\bibnamefont#1{#1}\fi
\expandafter\ifx\csname bibfnamefont\endcsname\relax
  \def\bibfnamefont#1{#1}\fi
\expandafter\ifx\csname citenamefont\endcsname\relax
  \def\citenamefont#1{#1}\fi
\expandafter\ifx\csname url\endcsname\relax
  \def\url#1{\texttt{#1}}\fi
\expandafter\ifx\csname urlprefix\endcsname\relax\def\urlprefix{URL }\fi
\providecommand{\bibinfo}[2]{#2}
\providecommand{\eprint}[2][]{\url{#2}}

\bibitem[{\citenamefont{Withers et~al.}(2015)\citenamefont{Withers, Del
  Pozo-Zamudio, Mishchenko, Rooney, Gholinia, Watanabe, Taniguchi, Haigh, Geim,
  Tartakovskii et~al.}}]{withers_light-emitting_2015}
\bibinfo{author}{\bibfnamefont{F.}~\bibnamefont{Withers}},
  \bibinfo{author}{\bibfnamefont{O.}~\bibnamefont{Del Pozo-Zamudio}},
  \bibinfo{author}{\bibfnamefont{A.}~\bibnamefont{Mishchenko}},
  \bibinfo{author}{\bibfnamefont{A.~P.} \bibnamefont{Rooney}},
  \bibinfo{author}{\bibfnamefont{A.}~\bibnamefont{Gholinia}},
  \bibinfo{author}{\bibfnamefont{K.}~\bibnamefont{Watanabe}},
  \bibinfo{author}{\bibfnamefont{T.}~\bibnamefont{Taniguchi}},
  \bibinfo{author}{\bibfnamefont{S.~J.} \bibnamefont{Haigh}},
  \bibinfo{author}{\bibfnamefont{A.~K.} \bibnamefont{Geim}},
  \bibinfo{author}{\bibfnamefont{A.~I.} \bibnamefont{Tartakovskii}},
  \bibnamefont{et~al.}, \bibinfo{journal}{Nature Materials}
  \textbf{\bibinfo{volume}{14}}, \bibinfo{pages}{301} (\bibinfo{year}{2015}),
  ISSN \bibinfo{issn}{1476-1122},
  \urlprefix\url{http://www.nature.com/nmat/journal/v14/n3/full/nmat4205.html}.

\bibitem[{\citenamefont{Wu et~al.}(2015)\citenamefont{Wu, Buckley, Schaibley,
  Feng, Yan, Mandrus, Hatami, Yao, Vučković, Majumdar
  et~al.}}]{wu_monolayer_2015}
\bibinfo{author}{\bibfnamefont{S.}~\bibnamefont{Wu}},
  \bibinfo{author}{\bibfnamefont{S.}~\bibnamefont{Buckley}},
  \bibinfo{author}{\bibfnamefont{J.~R.} \bibnamefont{Schaibley}},
  \bibinfo{author}{\bibfnamefont{L.}~\bibnamefont{Feng}},
  \bibinfo{author}{\bibfnamefont{J.}~\bibnamefont{Yan}},
  \bibinfo{author}{\bibfnamefont{D.~G.} \bibnamefont{Mandrus}},
  \bibinfo{author}{\bibfnamefont{F.}~\bibnamefont{Hatami}},
  \bibinfo{author}{\bibfnamefont{W.}~\bibnamefont{Yao}},
  \bibinfo{author}{\bibfnamefont{J.}~\bibnamefont{Vučković}},
  \bibinfo{author}{\bibfnamefont{A.}~\bibnamefont{Majumdar}},
  \bibnamefont{et~al.}, \bibinfo{journal}{Nature}
  \textbf{\bibinfo{volume}{520}}, \bibinfo{pages}{69} (\bibinfo{year}{2015}),
  ISSN \bibinfo{issn}{0028-0836},
  \urlprefix\url{http://www.nature.com/nature/journal/v520/n7545/full/nature14290.html}.

\bibitem[{\citenamefont{He et~al.}(2015)\citenamefont{He, Clark, Schaibley, He,
  Chen, Wei, Ding, Zhang, Yao, Xu et~al.}}]{he_single_2015}
\bibinfo{author}{\bibfnamefont{Y.-M.} \bibnamefont{He}},
  \bibinfo{author}{\bibfnamefont{G.}~\bibnamefont{Clark}},
  \bibinfo{author}{\bibfnamefont{J.~R.} \bibnamefont{Schaibley}},
  \bibinfo{author}{\bibfnamefont{Y.}~\bibnamefont{He}},
  \bibinfo{author}{\bibfnamefont{M.-C.} \bibnamefont{Chen}},
  \bibinfo{author}{\bibfnamefont{Y.-J.} \bibnamefont{Wei}},
  \bibinfo{author}{\bibfnamefont{X.}~\bibnamefont{Ding}},
  \bibinfo{author}{\bibfnamefont{Q.}~\bibnamefont{Zhang}},
  \bibinfo{author}{\bibfnamefont{W.}~\bibnamefont{Yao}},
  \bibinfo{author}{\bibfnamefont{X.}~\bibnamefont{Xu}}, \bibnamefont{et~al.},
  \bibinfo{journal}{Nature Nanotechnology} \textbf{\bibinfo{volume}{10}},
  \bibinfo{pages}{497} (\bibinfo{year}{2015}), ISSN \bibinfo{issn}{1748-3387},
  \urlprefix\url{http://www.nature.com/nnano/journal/v10/n6/full/nnano.2015.75.html}.

\bibitem[{\citenamefont{Chakraborty et~al.}(2015)\citenamefont{Chakraborty,
  Kinnischtzke, Goodfellow, Beams, and
  Vamivakas}}]{chakraborty_voltage-controlled_2015}
\bibinfo{author}{\bibfnamefont{C.}~\bibnamefont{Chakraborty}},
  \bibinfo{author}{\bibfnamefont{L.}~\bibnamefont{Kinnischtzke}},
  \bibinfo{author}{\bibfnamefont{K.~M.} \bibnamefont{Goodfellow}},
  \bibinfo{author}{\bibfnamefont{R.}~\bibnamefont{Beams}}, \bibnamefont{and}
  \bibinfo{author}{\bibfnamefont{A.~N.} \bibnamefont{Vamivakas}},
  \bibinfo{journal}{Nature Nanotechnology} \textbf{\bibinfo{volume}{10}},
  \bibinfo{pages}{507} (\bibinfo{year}{2015}), ISSN \bibinfo{issn}{1748-3387},
  \urlprefix\url{http://www.nature.com/nnano/journal/v10/n6/full/nnano.2015.79.html}.

\bibitem[{\citenamefont{Ugeda et~al.}(2014)\citenamefont{Ugeda, Bradley, Shi,
  da~Jornada, Zhang, Qiu, Ruan, Mo, Hussain, Shen et~al.}}]{ugeda_giant_2014}
\bibinfo{author}{\bibfnamefont{M.~M.} \bibnamefont{Ugeda}},
  \bibinfo{author}{\bibfnamefont{A.~J.} \bibnamefont{Bradley}},
  \bibinfo{author}{\bibfnamefont{S.-F.} \bibnamefont{Shi}},
  \bibinfo{author}{\bibfnamefont{F.~H.} \bibnamefont{da~Jornada}},
  \bibinfo{author}{\bibfnamefont{Y.}~\bibnamefont{Zhang}},
  \bibinfo{author}{\bibfnamefont{D.~Y.} \bibnamefont{Qiu}},
  \bibinfo{author}{\bibfnamefont{W.}~\bibnamefont{Ruan}},
  \bibinfo{author}{\bibfnamefont{S.-K.} \bibnamefont{Mo}},
  \bibinfo{author}{\bibfnamefont{Z.}~\bibnamefont{Hussain}},
  \bibinfo{author}{\bibfnamefont{Z.-X.} \bibnamefont{Shen}},
  \bibnamefont{et~al.}, \bibinfo{journal}{Nature Materials}
  \textbf{\bibinfo{volume}{13}}, \bibinfo{pages}{1091} (\bibinfo{year}{2014}),
  ISSN \bibinfo{issn}{1476-1122},
  \urlprefix\url{http://www.nature.com/nmat/journal/v13/n12/full/nmat4061.html}.

\bibitem[{\citenamefont{Poellmann et~al.}(2015)\citenamefont{Poellmann,
  Steinleitner, Leierseder, Nagler, Plechinger, Porer, Bratschitsch, Schüller,
  Korn, and Huber}}]{poellmann_resonant_2015}
\bibinfo{author}{\bibfnamefont{C.}~\bibnamefont{Poellmann}},
  \bibinfo{author}{\bibfnamefont{P.}~\bibnamefont{Steinleitner}},
  \bibinfo{author}{\bibfnamefont{U.}~\bibnamefont{Leierseder}},
  \bibinfo{author}{\bibfnamefont{P.}~\bibnamefont{Nagler}},
  \bibinfo{author}{\bibfnamefont{G.}~\bibnamefont{Plechinger}},
  \bibinfo{author}{\bibfnamefont{M.}~\bibnamefont{Porer}},
  \bibinfo{author}{\bibfnamefont{R.}~\bibnamefont{Bratschitsch}},
  \bibinfo{author}{\bibfnamefont{C.}~\bibnamefont{Schüller}},
  \bibinfo{author}{\bibfnamefont{T.}~\bibnamefont{Korn}}, \bibnamefont{and}
  \bibinfo{author}{\bibfnamefont{R.}~\bibnamefont{Huber}},
  \bibinfo{journal}{Nature Materials} \textbf{\bibinfo{volume}{14}},
  \bibinfo{pages}{889} (\bibinfo{year}{2015}), ISSN \bibinfo{issn}{1476-1122},
  \urlprefix\url{http://www.nature.com/nmat/journal/v14/n9/abs/nmat4356.html}.

\bibitem[{\citenamefont{Nie et~al.}(2014)\citenamefont{Nie, Long, Sun, Huang,
  Zhang, Xiong, Hewak, Shen, Prezhdo, and Loh}}]{nie_ultrafast_2014}
\bibinfo{author}{\bibfnamefont{Z.}~\bibnamefont{Nie}},
  \bibinfo{author}{\bibfnamefont{R.}~\bibnamefont{Long}},
  \bibinfo{author}{\bibfnamefont{L.}~\bibnamefont{Sun}},
  \bibinfo{author}{\bibfnamefont{C.-C.} \bibnamefont{Huang}},
  \bibinfo{author}{\bibfnamefont{J.}~\bibnamefont{Zhang}},
  \bibinfo{author}{\bibfnamefont{Q.}~\bibnamefont{Xiong}},
  \bibinfo{author}{\bibfnamefont{D.~W.} \bibnamefont{Hewak}},
  \bibinfo{author}{\bibfnamefont{Z.}~\bibnamefont{Shen}},
  \bibinfo{author}{\bibfnamefont{O.~V.} \bibnamefont{Prezhdo}},
  \bibnamefont{and} \bibinfo{author}{\bibfnamefont{Z.-H.} \bibnamefont{Loh}},
  \bibinfo{journal}{{ACS} Nano} \textbf{\bibinfo{volume}{8}},
  \bibinfo{pages}{10931} (\bibinfo{year}{2014}), ISSN
  \bibinfo{issn}{1936-0851},
  \urlprefix\url{http://dx.doi.org/10.1021/nn504760x}.

\bibitem[{\citenamefont{Chernikov et~al.}(2015)\citenamefont{Chernikov,
  Ruppert, Hill, Rigosi, and Heinz}}]{chernikov_population_2015}
\bibinfo{author}{\bibfnamefont{A.}~\bibnamefont{Chernikov}},
  \bibinfo{author}{\bibfnamefont{C.}~\bibnamefont{Ruppert}},
  \bibinfo{author}{\bibfnamefont{H.~M.} \bibnamefont{Hill}},
  \bibinfo{author}{\bibfnamefont{A.~F.} \bibnamefont{Rigosi}},
  \bibnamefont{and} \bibinfo{author}{\bibfnamefont{T.~F.} \bibnamefont{Heinz}},
  \bibinfo{journal}{Nature Photonics} \textbf{\bibinfo{volume}{9}},
  \bibinfo{pages}{466} (\bibinfo{year}{2015}), ISSN \bibinfo{issn}{1749-4885,
  1749-4893},
  \urlprefix\url{http://www.nature.com/doifinder/10.1038/nphoton.2015.104}.

\bibitem[{\citenamefont{Moody et~al.}(2015)\citenamefont{Moody, Kavir~Dass,
  Hao, Chen, Li, Singh, Tran, Clark, Xu, Berghäuser
  et~al.}}]{moody_intrinsic_2015}
\bibinfo{author}{\bibfnamefont{G.}~\bibnamefont{Moody}},
  \bibinfo{author}{\bibfnamefont{C.}~\bibnamefont{Kavir~Dass}},
  \bibinfo{author}{\bibfnamefont{K.}~\bibnamefont{Hao}},
  \bibinfo{author}{\bibfnamefont{C.-H.} \bibnamefont{Chen}},
  \bibinfo{author}{\bibfnamefont{L.-J.} \bibnamefont{Li}},
  \bibinfo{author}{\bibfnamefont{A.}~\bibnamefont{Singh}},
  \bibinfo{author}{\bibfnamefont{K.}~\bibnamefont{Tran}},
  \bibinfo{author}{\bibfnamefont{G.}~\bibnamefont{Clark}},
  \bibinfo{author}{\bibfnamefont{X.}~\bibnamefont{Xu}},
  \bibinfo{author}{\bibfnamefont{G.}~\bibnamefont{Berghäuser}},
  \bibnamefont{et~al.}, \bibinfo{journal}{Nature Communications}
  \textbf{\bibinfo{volume}{6}}, \bibinfo{pages}{8315} (\bibinfo{year}{2015}),
  \urlprefix\url{http://www.nature.com/ncomms/2015/150918/ncomms9315/full/ncomms9315.html}.

\bibitem[{\citenamefont{Steinhoff et~al.}(2014)\citenamefont{Steinhoff,
  R\"osner, Jahnke, Wehling, and Gies}}]{steinhoff_influence_2014}
\bibinfo{author}{\bibfnamefont{A.}~\bibnamefont{Steinhoff}},
  \bibinfo{author}{\bibfnamefont{M.}~\bibnamefont{R\"osner}},
  \bibinfo{author}{\bibfnamefont{F.}~\bibnamefont{Jahnke}},
  \bibinfo{author}{\bibfnamefont{T.~O.} \bibnamefont{Wehling}},
  \bibnamefont{and} \bibinfo{author}{\bibfnamefont{C.}~\bibnamefont{Gies}},
  \bibinfo{journal}{Nano Letters} \textbf{\bibinfo{volume}{14}},
  \bibinfo{pages}{3743} (\bibinfo{year}{2014}), ISSN \bibinfo{issn}{1530-6984},
  \urlprefix\url{http://dx.doi.org/10.1021/nl500595u}.

\bibitem[{\citenamefont{Hill et~al.}(2015)\citenamefont{Hill, Rigosi, Roquelet,
  Chernikov, Berkelbach, Reichman, Hybertsen, Brus, and
  Heinz}}]{hill_observation_2015}
\bibinfo{author}{\bibfnamefont{H.~M.} \bibnamefont{Hill}},
  \bibinfo{author}{\bibfnamefont{A.~F.} \bibnamefont{Rigosi}},
  \bibinfo{author}{\bibfnamefont{C.}~\bibnamefont{Roquelet}},
  \bibinfo{author}{\bibfnamefont{A.}~\bibnamefont{Chernikov}},
  \bibinfo{author}{\bibfnamefont{T.~C.} \bibnamefont{Berkelbach}},
  \bibinfo{author}{\bibfnamefont{D.~R.} \bibnamefont{Reichman}},
  \bibinfo{author}{\bibfnamefont{M.~S.} \bibnamefont{Hybertsen}},
  \bibinfo{author}{\bibfnamefont{L.~E.} \bibnamefont{Brus}}, \bibnamefont{and}
  \bibinfo{author}{\bibfnamefont{T.~F.} \bibnamefont{Heinz}},
  \bibinfo{journal}{Nano Letters} \textbf{\bibinfo{volume}{15}},
  \bibinfo{pages}{2992} (\bibinfo{year}{2015}), ISSN \bibinfo{issn}{1530-6984},
  \urlprefix\url{http://dx.doi.org/10.1021/nl504868p}.

\bibitem[{\citenamefont{Qiu et~al.}(2013)\citenamefont{Qiu, da~Jornada, and
  Louie}}]{qiu_optical_2013}
\bibinfo{author}{\bibfnamefont{D.~Y.} \bibnamefont{Qiu}},
  \bibinfo{author}{\bibfnamefont{F.~H.} \bibnamefont{da~Jornada}},
  \bibnamefont{and} \bibinfo{author}{\bibfnamefont{S.~G.} \bibnamefont{Louie}},
  \bibinfo{journal}{Physical Review Letters} \textbf{\bibinfo{volume}{111}},
  \bibinfo{pages}{216805} (\bibinfo{year}{2013}),
  \urlprefix\url{http://link.aps.org/doi/10.1103/PhysRevLett.111.216805}.

\bibitem[{\citenamefont{Mai et~al.}(2014)\citenamefont{Mai, Barrette, Yu,
  Semenov, Kim, Cao, and Gundogdu}}]{mai_many-body_2014}
\bibinfo{author}{\bibfnamefont{C.}~\bibnamefont{Mai}},
  \bibinfo{author}{\bibfnamefont{A.}~\bibnamefont{Barrette}},
  \bibinfo{author}{\bibfnamefont{Y.}~\bibnamefont{Yu}},
  \bibinfo{author}{\bibfnamefont{Y.~G.} \bibnamefont{Semenov}},
  \bibinfo{author}{\bibfnamefont{K.~W.} \bibnamefont{Kim}},
  \bibinfo{author}{\bibfnamefont{L.}~\bibnamefont{Cao}}, \bibnamefont{and}
  \bibinfo{author}{\bibfnamefont{K.}~\bibnamefont{Gundogdu}},
  \bibinfo{journal}{Nano Letters} \textbf{\bibinfo{volume}{14}},
  \bibinfo{pages}{202} (\bibinfo{year}{2014}), ISSN \bibinfo{issn}{1530-6984},
  \urlprefix\url{http://dx.doi.org/10.1021/nl403742j}.

\bibitem[{\citenamefont{Yang et~al.}(2015)\citenamefont{Yang, Sinitsyn, Chen,
  Yuan, Zhang, Lou, and Crooker}}]{yang_long-lived_2015}
\bibinfo{author}{\bibfnamefont{L.}~\bibnamefont{Yang}},
  \bibinfo{author}{\bibfnamefont{N.~A.} \bibnamefont{Sinitsyn}},
  \bibinfo{author}{\bibfnamefont{W.}~\bibnamefont{Chen}},
  \bibinfo{author}{\bibfnamefont{J.}~\bibnamefont{Yuan}},
  \bibinfo{author}{\bibfnamefont{J.}~\bibnamefont{Zhang}},
  \bibinfo{author}{\bibfnamefont{J.}~\bibnamefont{Lou}}, \bibnamefont{and}
  \bibinfo{author}{\bibfnamefont{S.~A.} \bibnamefont{Crooker}},
  \bibinfo{journal}{Nature Physics} \textbf{\bibinfo{volume}{11}},
  \bibinfo{pages}{830} (\bibinfo{year}{2015}), ISSN \bibinfo{issn}{1745-2473},
  \urlprefix\url{http://www.nature.com/nphys/journal/v11/n10/full/nphys3419.html}.

\bibitem[{\citenamefont{Malic et~al.}(2011)\citenamefont{Malic, Winzer, Bobkin,
  and Knorr}}]{malic_microscopic_2011}
\bibinfo{author}{\bibfnamefont{E.}~\bibnamefont{Malic}},
  \bibinfo{author}{\bibfnamefont{T.}~\bibnamefont{Winzer}},
  \bibinfo{author}{\bibfnamefont{E.}~\bibnamefont{Bobkin}}, \bibnamefont{and}
  \bibinfo{author}{\bibfnamefont{A.}~\bibnamefont{Knorr}},
  \bibinfo{journal}{Physical Review B} \textbf{\bibinfo{volume}{84}},
  \bibinfo{pages}{205406} (\bibinfo{year}{2011}),
  \urlprefix\url{http://link.aps.org/doi/10.1103/PhysRevB.84.205406}.

\bibitem[{\citenamefont{Conley et~al.}(2013)\citenamefont{Conley, Wang,
  Ziegler, Haglund, Pantelides, and Bolotin}}]{conley_bandgap_2013}
\bibinfo{author}{\bibfnamefont{H.~J.} \bibnamefont{Conley}},
  \bibinfo{author}{\bibfnamefont{B.}~\bibnamefont{Wang}},
  \bibinfo{author}{\bibfnamefont{J.~I.} \bibnamefont{Ziegler}},
  \bibinfo{author}{\bibfnamefont{R.~F.} \bibnamefont{Haglund}},
  \bibinfo{author}{\bibfnamefont{S.~T.} \bibnamefont{Pantelides}},
  \bibnamefont{and} \bibinfo{author}{\bibfnamefont{K.~I.}
  \bibnamefont{Bolotin}}, \bibinfo{journal}{Nano Letters}
  \textbf{\bibinfo{volume}{13}}, \bibinfo{pages}{3626} (\bibinfo{year}{2013}),
  ISSN \bibinfo{issn}{1530-6984},
  \urlprefix\url{http://dx.doi.org/10.1021/nl4014748}.

\bibitem[{\citenamefont{Steinhoff et~al.}(2015)\citenamefont{Steinhoff, Kim,
  Jahnke, R\"osner, Kim, Lee, Han, Jeong, Wehling, and
  Gies}}]{steinhoff_efficient_2015}
\bibinfo{author}{\bibfnamefont{A.}~\bibnamefont{Steinhoff}},
  \bibinfo{author}{\bibfnamefont{J.-H.} \bibnamefont{Kim}},
  \bibinfo{author}{\bibfnamefont{F.}~\bibnamefont{Jahnke}},
  \bibinfo{author}{\bibfnamefont{M.}~\bibnamefont{R\"osner}},
  \bibinfo{author}{\bibfnamefont{D.-S.} \bibnamefont{Kim}},
  \bibinfo{author}{\bibfnamefont{C.}~\bibnamefont{Lee}},
  \bibinfo{author}{\bibfnamefont{G.~H.} \bibnamefont{Han}},
  \bibinfo{author}{\bibfnamefont{M.~S.} \bibnamefont{Jeong}},
  \bibinfo{author}{\bibfnamefont{T.~O.} \bibnamefont{Wehling}},
  \bibnamefont{and} \bibinfo{author}{\bibfnamefont{C.}~\bibnamefont{Gies}},
  \bibinfo{journal}{Nano Letters} \textbf{\bibinfo{volume}{15}},
  \bibinfo{pages}{6841} (\bibinfo{year}{2015}), ISSN \bibinfo{issn}{1530-6984},
  \urlprefix\url{http://dx.doi.org/10.1021/acs.nanolett.5b02719}.

\bibitem[{\citenamefont{Knox et~al.}(1988)\citenamefont{Knox, Chemla, Livescu,
  Cunningham, and Henry}}]{knox_femtosecond_1988}
\bibinfo{author}{\bibfnamefont{W.~H.} \bibnamefont{Knox}},
  \bibinfo{author}{\bibfnamefont{D.~S.} \bibnamefont{Chemla}},
  \bibinfo{author}{\bibfnamefont{G.}~\bibnamefont{Livescu}},
  \bibinfo{author}{\bibfnamefont{J.~E.} \bibnamefont{Cunningham}},
  \bibnamefont{and} \bibinfo{author}{\bibfnamefont{J.~E.} \bibnamefont{Henry}},
  \bibinfo{journal}{Physical Review Letters} \textbf{\bibinfo{volume}{61}},
  \bibinfo{pages}{1290} (\bibinfo{year}{1988}),
  \urlprefix\url{http://link.aps.org/doi/10.1103/PhysRevLett.61.1290}.

\bibitem[{\citenamefont{Knox et~al.}(1986)\citenamefont{Knox, Hirlimann,
  Miller, Shah, Chemla, and Shank}}]{knox_femtosecond_1986}
\bibinfo{author}{\bibfnamefont{W.~H.} \bibnamefont{Knox}},
  \bibinfo{author}{\bibfnamefont{C.}~\bibnamefont{Hirlimann}},
  \bibinfo{author}{\bibfnamefont{D.~A.~B.} \bibnamefont{Miller}},
  \bibinfo{author}{\bibfnamefont{J.}~\bibnamefont{Shah}},
  \bibinfo{author}{\bibfnamefont{D.~S.} \bibnamefont{Chemla}},
  \bibnamefont{and} \bibinfo{author}{\bibfnamefont{C.~V.} \bibnamefont{Shank}},
  \bibinfo{journal}{Physical Review Letters} \textbf{\bibinfo{volume}{56}},
  \bibinfo{pages}{1191} (\bibinfo{year}{1986}),
  \urlprefix\url{http://link.aps.org/doi/10.1103/PhysRevLett.56.1191}.

\bibitem[{\citenamefont{Shah}(1999)}]{shah_ultrafast_1999}
\bibinfo{author}{\bibfnamefont{J.}~\bibnamefont{Shah}},
  \emph{\bibinfo{title}{Ultrafast Spectroscopy of Semiconductors and
  Semiconductor Nanostructures}}, vol. \bibinfo{volume}{115} of
  \emph{\bibinfo{series}{Springer Series in Solid-State Sciences}}
  (\bibinfo{publisher}{Springer Berlin Heidelberg}, \bibinfo{address}{Berlin,
  Heidelberg}, \bibinfo{year}{1999}), ISBN \bibinfo{isbn}{978-3-642-08391-4
  978-3-662-03770-6},
  \urlprefix\url{http://link.springer.com/10.1007/978-3-662-03770-6}.

\bibitem[{\citenamefont{Sch\"afer and
  Wegener}(2002)}]{schafer_semiconductor_2002}
\bibinfo{author}{\bibfnamefont{W.}~\bibnamefont{Sch\"afer}} \bibnamefont{and}
  \bibinfo{author}{\bibfnamefont{M.}~\bibnamefont{Wegener}},
  \emph{\bibinfo{title}{Semiconductor optics and transport phenomena}}
  (\bibinfo{publisher}{Springer}, \bibinfo{address}{Berlin; New York},
  \bibinfo{year}{2002}), ISBN \bibinfo{isbn}{3-540-61614-4 978-3-540-61614-6
  3-642-08271-8 978-3-642-08271-9}.

\bibitem[{\citenamefont{Liu et~al.}(2013)\citenamefont{Liu, Shan, Yao, Yao, and
  Xiao}}]{liu_three-band_2013}
\bibinfo{author}{\bibfnamefont{G.-B.} \bibnamefont{Liu}},
  \bibinfo{author}{\bibfnamefont{W.-Y.} \bibnamefont{Shan}},
  \bibinfo{author}{\bibfnamefont{Y.}~\bibnamefont{Yao}},
  \bibinfo{author}{\bibfnamefont{W.}~\bibnamefont{Yao}}, \bibnamefont{and}
  \bibinfo{author}{\bibfnamefont{D.}~\bibnamefont{Xiao}},
  \bibinfo{journal}{Physical Review B} \textbf{\bibinfo{volume}{88}},
  \bibinfo{pages}{085433} (\bibinfo{year}{2013}),
  \urlprefix\url{http://link.aps.org/doi/10.1103/PhysRevB.88.085433}.

\bibitem[{\citenamefont{Haug and Koch}(1993)}]{haug_quantum_1993}
\bibinfo{author}{\bibfnamefont{H.}~\bibnamefont{Haug}} \bibnamefont{and}
  \bibinfo{author}{\bibfnamefont{S.~W.} \bibnamefont{Koch}},
  \emph{\bibinfo{title}{Quantum theory of the optical and electronic properties
  of semiconductors}} (\bibinfo{publisher}{World Scientific},
  \bibinfo{address}{Singapore; River Edge, {NJ}}, \bibinfo{year}{1993}), ISBN
  \bibinfo{isbn}{981-02-1341-7 978-981-02-1341-1 981-02-1347-6
  978-981-02-1347-3}.

\bibitem[{\citenamefont{Nie et~al.}(2015)\citenamefont{Nie, Long, Teguh, Huang,
  Hewak, Yeow, Shen, Prezhdo, and Loh}}]{nie_ultrafast_2015}
\bibinfo{author}{\bibfnamefont{Z.}~\bibnamefont{Nie}},
  \bibinfo{author}{\bibfnamefont{R.}~\bibnamefont{Long}},
  \bibinfo{author}{\bibfnamefont{J.~S.} \bibnamefont{Teguh}},
  \bibinfo{author}{\bibfnamefont{C.-C.} \bibnamefont{Huang}},
  \bibinfo{author}{\bibfnamefont{D.~W.} \bibnamefont{Hewak}},
  \bibinfo{author}{\bibfnamefont{E.~K.~L.} \bibnamefont{Yeow}},
  \bibinfo{author}{\bibfnamefont{Z.}~\bibnamefont{Shen}},
  \bibinfo{author}{\bibfnamefont{O.~V.} \bibnamefont{Prezhdo}},
  \bibnamefont{and} \bibinfo{author}{\bibfnamefont{Z.-H.} \bibnamefont{Loh}},
  \bibinfo{journal}{The Journal of Physical Chemistry C}
  \textbf{\bibinfo{volume}{119}}, \bibinfo{pages}{20698}
  (\bibinfo{year}{2015}), ISSN \bibinfo{issn}{1932-7447},
  \urlprefix\url{http://dx.doi.org/10.1021/acs.jpcc.5b05048}.

\bibitem[{\citenamefont{Berntsen and Espelid}(1992)}]{berntsen_algorithm_1992}
\bibinfo{author}{\bibfnamefont{J.}~\bibnamefont{Berntsen}} \bibnamefont{and}
  \bibinfo{author}{\bibfnamefont{T.~O.} \bibnamefont{Espelid}},
  \bibinfo{journal}{{ACM} Trans. Math. Softw.} \textbf{\bibinfo{volume}{18}},
  \bibinfo{pages}{329–342} (\bibinfo{year}{1992}), ISSN
  \bibinfo{issn}{0098-3500},
  \urlprefix\url{http://doi.acm.org/10.1145/131766.131772}.

\bibitem[{\citenamefont{Renka}(1999)}]{renka_algorithm_1999}
\bibinfo{author}{\bibfnamefont{R.~J.} \bibnamefont{Renka}},
  \bibinfo{journal}{{ACM} Trans. Math. Softw.} \textbf{\bibinfo{volume}{25}},
  \bibinfo{pages}{70–73} (\bibinfo{year}{1999}), ISSN
  \bibinfo{issn}{0098-3500},
  \urlprefix\url{http://doi.acm.org/10.1145/305658.305737}.

\bibitem[{\citenamefont{Brown et~al.}(1989)\citenamefont{Brown, Byrne, and
  Hindmarsh}}]{brown_vode:_1989}
\bibinfo{author}{\bibfnamefont{P.}~\bibnamefont{Brown}},
  \bibinfo{author}{\bibfnamefont{G.}~\bibnamefont{Byrne}}, \bibnamefont{and}
  \bibinfo{author}{\bibfnamefont{A.}~\bibnamefont{Hindmarsh}},
  \bibinfo{journal}{{SIAM} Journal on Scientific and Statistical Computing}
  \textbf{\bibinfo{volume}{10}}, \bibinfo{pages}{1038} (\bibinfo{year}{1989}),
  ISSN \bibinfo{issn}{0196-5204},
  \urlprefix\url{http://epubs.siam.org/doi/abs/10.1137/0910062}.

\end{thebibliography}

\end{document}